\documentstyle[11pt,newpasp,twoside]{article}
\def\grtsim{{_ >\atop{^\sim}}}

\def\edcomment#1{\iffalse\marginpar{\raggedright\sl#1\/}\else\relax\fi}
\reversemarginpar

\begin{document}
\title{Properties and evolution of disk galaxies in a
hierarchical formation scenario}
\author{Vladimir Avila-Reese and Claudio Firmani\altaffilmark{1}}
\affil{Instituto de Astronom\'\i a-UNAM, A.P. 70-264, 04510 
M\'exico, D. F.}

\altaffiltext{1}{Also Osservatorio Astronomico di Brera, via E.Bianchi 
46, I-23807 Merate, Italy}

\begin{abstract}
We highlight some results from disk galaxy evolution models conceived 
within a cosmological context. When disk mergers and strong disk-halo
feedback are omitted, several properties and correlations
of disk galaxies seem to be related to initial conditions given
by the CDM model.  
\end{abstract}

$\bullet${\bf Motivation.} The inflation-inspired cold dark matter (CDM) 
hierarchical model has provided an invaluable 
theoretical framework for studies on galaxy formation and 
evolution. Nevertheless, several aspects of this phenomenon can 
not be treated as a simple (deductive) extrapolation of the hierarchical
scenario, as is commonly done in some approaches. We have developed
an approach where an inductive (backward) disk galaxy evolution
model is combined with initial and boundary conditions calculated 
from the hierarchical scenario. Our main goal is {\it to find the relevant 
initial factors and physical ingredients that determine
the main local and global properties, correlations and evolutionary 
features of disk galaxies.}    
In our approach (Avila-Reese et al. 1998; Avila-Reese \& Firmani 2000a;
Firmani \& Avila-Reese 2000), disks form inside-out within growing CDM halos 
with a gentle gas accretion rate (no mergers) proportional to the 
hierarchical mass aggregation rate. We follow locally the overall evolution 
of individual disks in centrifugal equilibrium, including self-regulated
star formation (SF) by feedback {\it within the disk ISM} (no disk-halo
feedback) and population synthesis. Bulges are assumed to form
via a secular mechanism. At the same time, as in the semianalytical 
approach, we are able to predict correlations and statistical 
properties of the disk galaxy population. 

$\bullet$ {\bf The disk Hubble sequence.} We 
find that the main properties and correlations of disk galaxy 
models are mainly determined by the 
cosmological conditions. For a given mass, the dark halo 
concentration, the galaxy color index and the disk gas fraction 
are basically related to the {\bf mass aggregation history (MAH)}, while
the disk surface brightness (SB), the bulge-to-disk ratio (b/d), and the 
shape of the rotation curve depend mainly on the {\bf spin 
parameter $\lambda$}. The mass does not influence intensive
properties. Thus, according to our models, the {\it Hubble sequence
is biparametrical}, the MAH and $\lambda$ being the two driving
physical factors. The observational trends across the Hubble 
sequence are indeed reproduced: the redder and more concentrated
is the disk, the smaller is the gas fraction and the larger is 
the b/d ratio. The disk mass 
fraction $f_d$ also influences the models; we constrict
its value to $0.03<f_d<0.08$, otherwise the Tully-Fisher relation
(TFR) of LSB and HSB galaxies would be very different (lower limit) 
or the rotation curves would be too peaked (upper limit).
 
$\bullet$ {\bf Are the infrared TFRs cosmological or not?} 
For the popular flat $\Omega_{\Lambda}=h=0.7$ CDM cosmology,
the predicted slope and zero-point of the infrared TFRs is within 
the range of observational determinations, the
agreement being even better when shallow cores are introduced
in the CDM halos. In our models and for the mentioned cosmology, 
the maximum circular velocity increases on average $20-25\%$ after 
disk formation which is $\sim 2$ times less than Steinmetz \& Navarro
(1999) reported from their N-body+hydrodynamic simulations (the ``angular
momentum catastrophe'' problem). This factor of 2 in velocity
translates into a factor of $\sim 8$ in mass or luminosity explaining
why these authors obtain a TFR zero-point 2 magnitudes fainter than
observed. The slope of our modeled TFR ($\sim 3.4$) is essentially
the slope of the halo mass-velocity relation (variations in the 
disk mass fraction only scarcely affect this slope). Since observations
seem to show finally an upper limit of $\sim 3.4$ for the slope
of the most infrared TFR (Tully \& Pierce 1999; a similar slope was
also found for the baryonic TFR, de Jong \& Bell, this volume), 
our result suggest that there is not room for intermediate astrophysical 
ingredients like a dependence of SF efficiency or halo gas reheating on 
mass. Regarding the TFR scatter, we find it to be between 0.38 and 0.31 mag
for velocities ranking from 70 to 300 km/s.    

We also find that the {\bf LSB and HSB galaxies have roughly the 
same TFR:} although for a given mass the $V_{\rm max}$ of 
LSB galaxies is less than that of HSB galaxies, the disk luminosity
is also less since the SF efficiency depends on disk surface density. 
This also explains why the residuals of the TFR do not correlate 
strongly with the disk lenghtscale (or SB), avoiding this way the 
interpretation of Courteau \& Rix (1997) of large amounts of dark 
matter in the inner parts of disk galaxies which is at odds with 
observations showing the shape of the rotation curves to correlate
with SB for a given luminosity. 
      
$\bullet${\bf Some evolutionary features} (see Avila-Reese \& Firmani 2000b).
 The shape of the SF history (SFH) in our
models depends on the MAH and $\lambda$ (SB). For most of the cases,
this shape has a broad maximum at $z\approx 1.5-2.5$ and a fall towards
$z=0$ by factors 2-4. This factor is $\sim 6-10$ in the measured
{\it global} SFH. Then, other galaxy populations than disk galaxies had to
have contributed in the past to the global SF rate. The $B-$band 
luminosity is larger
and the galaxy colors become slightly bluer in the past. The disk sizes
strongly decrease towards the past (a factor of 2 at $z\grtsim1$ w.r.t.
$z=0$), while the SB increases.  

V.A. received finantial support from CONACyT grant 27752-E.

\end{document}